\documentclass[letterpaper,11pt]{article}
\pdfoutput=1 

\usepackage{jheppub} 

\usepackage{graphicx}
\usepackage{caption}
\usepackage{subcaption}
\usepackage{multirow}
\usepackage{amssymb}
\usepackage{epstopdf}
\usepackage{verbatim}
\usepackage{slashed}

\usepackage{url}
\newcommand{\be}{\begin{equation}}
\newcommand{\bea}{\begin{eqnarray}}
\newcommand{\beq}[1]{\begin{equation}\label{#1}}

\newcommand{\ee}{\end{equation}}
\newcommand{\eea}{\end{eqnarray}}
\newcommand{\eeq}{\end{equation}}
\newcommand{\lsim}{\!\mathrel{\hbox{\rlap{\lower.55ex \hbox{$\sim$}} \kern-.34em \raise.4ex \hbox{$<$}}}}
\newcommand{\gsim}{\!\mathrel{\hbox{\rlap{\lower.55ex \hbox{$\sim$}} \kern-.34em \raise.4ex \hbox{$>$}}}}

\newcommand{\abs}[1]{\left| #1 \right|}

\newcommand{\Sec}[1]{Sec.~\ref{#1}}

\newcommand{\Tab}[1]{Table~\ref{#1}}

\newcommand{\Fig}[1]{Fig.~\ref{#1}}

\newcommand{\Eq}[1]{Eq.~(\ref{#1})}

\newcommand{\Ref}[1]{Ref.~\cite{#1}}
\newcommand{\Refs}[1]{Refs.~\cite{#1}}

\begin{document}

\begin{flushright}MCTP-13-11 \\
MIT-CTP 4453
\end{flushright}
\vspace{0.2cm}

\title{Top Partner Probes of Extended Higgs Sectors}

\author[a,1]{John Kearney,\note{Corresponding author.}}
\author[a]{Aaron Pierce,}
\author[b]{and Jesse Thaler}
\affiliation[a]{Michigan Center for Theoretical Physics, Department of Physics, Ann Arbor, MI 48109, USA}
\affiliation[b]{Center for Theoretical Physics, Massachusetts Institute of Technology, Cambridge, MA 02139, USA}

\emailAdd{jkrny@umich.edu}
\emailAdd{atpierce@umich.edu}
\emailAdd{jthaler@mit.edu}

\abstract{Natural theories of the weak scale often include fermionic partners of the top quark.  If the electroweak symmetry breaking sector contains scalars beyond a single Higgs doublet, then top partners can have sizable branching ratios to these extended Higgs sector states.  In fact, top partner decays may provide the most promising discovery mode for such scalars, especially given the large backgrounds to direct and associated production.
In this paper, we present a search strategy for top partner decays to a charged Higgs boson and a bottom quark, focusing on the case where the charged Higgs dominantly decays to third-generation quarks to yield a multi-$b$ final state.  We also discuss ways to extend this search to exotic neutral scalars decaying to bottom quark pairs.}

\maketitle

\section{Introduction}

With the discovery of a 125 GeV Higgs boson at the LHC, we are getting our first glimpse at the origin of electroweak symmetry breaking (EWSB). If naturalness is a reliable guide, then we expect additional dynamics at the TeV scale to regulate quadratic divergences in the Higgs potential.  In the Standard Model (SM), the large top quark Yukawa coupling induces large radiative corrections to $m_h^2$.  Consequently, models of new physics generally involve new colored particles, top partners, to cancel these quadratic divergences.  In the case where the Higgs is a pseudo-Nambu-Goldstone boson (PNGB), such top partners are fermionic.  

Another common feature of new physics models is an extended Higgs sector, which often involves a second Higgs doublet or additional singlet scalars.  This feature is particularly prevalent when the Higgs arises as a PNGB, since the breaking of a global symmetry $G \rightarrow H$ often gives rise to more than just a single complex Higgs doublet.  Because these extended scalars states typically carry only electroweak quantum numbers, they have small direct production cross sections at hadron colliders like the LHC.  Therefore, it is important to explore new search strategies in order to fully investigate the possible dynamics of EWSB.

In this paper, we show how top partners can open additional discovery channels for extended Higgs sectors.  In particular, top partners can be copiously pair-produced at the LHC through QCD processes, and the decay of top partners may provide the best avenue for observing additional scalars.  For concreteness, we will focus on the decay of a top partner $T$ to a charged Higgs $H^\pm$ and a bottom quark $b$,
\be
T \to b H^\pm, \qquad H^\pm \to t b, 
\ee
where we utilize the charged Higgs decay mode that typically dominates for $m_{H^\pm} > m_t + m_b$.  We will also show how the same search strategy is sensitive to neutral singlets $\varphi^0$ via
\be
T \to t \varphi^0, \qquad  \varphi^0 \to b b.
\ee
However, we wish to emphasize a more general point: \emph{if new top partners are found, searches for exotic decays to scalars should be a priority}. Our approach shares some intellectual ancestry with strategies to find Higgs bosons through supersymmetric particle decays \cite{Datta:2001qs,Datta:2003iz}, as well as studies designed to pick out the SM Higgs boson from top partner decays using jet substructure techniques \cite{Kribs:2010ii}.

Previous studies of the detectability of charged Higgs states with $m_{H^\pm} > m_t + m_b$ have focused on top quark associated production $g b \rightarrow t H^\pm$ \cite{Barger:1993th, Gunion:1993sv}.  The cross section for this process can in principle be large because extended Higgs sector states often have significant couplings to top quarks.  However, as we will review, there are a number of obstacles that make this search challenging.  Assuming top partners exist, we will show how pair production of top partners followed by the decay $T \to b H^\pm$ can be a complementary search strategy.  Should these exotic top partner decays be observed, they will become an important window  to the structure of new physics at the TeV scale.

The proposed search is particularly well-motivated by little Higgs (LH) scenarios \cite{ArkaniHamed:2001nc,ArkaniHamed:2002qx,ArkaniHamed:2002qy,Schmaltz:2005ky,Perelstein:2005ka}, which prominently feature both top partners and extended Higgs sectors \cite{Kearney:2013cca}.  In fact, LH models often contain more top partners than strictly necessary to regulate the Higgs potential, perhaps because of an underlying custodial symmetry \cite{Chang:2003un,Chang:2003zn,Agashe:2006at} or an enhanced global symmetry of the strong dynamics \cite{Katz:2003sn,Thaler:2005en}.  The search described here is relevant for standard top partners as well as their exotic cousins.  Similarly, as emphasized in \Ref{Schmaltz:2008vd}, the scalar sector of LH models must contain more than just a single Higgs doublet.  At minimum, additional scalars are necessary to achieve the desired the quartic potential for the Higgs boson.  Moreover, unless the theory has a symmetry like $T$-parity \cite{Cheng:2003ju,Cheng:2004yc}, precision electroweak constraints plus the model building constraint of ``dangerous singlets'' imply the presence of at least two Higgs doublets \cite{Schmaltz:2008vd}.  While we are motivated by LH models, the phenomenology we discuss in this paper is relevant for any theory with exotic top-like states and extended Higgs sectors.  For example, similar phenomenology can be present in heavy fourth generation models with multiple Higgs doublets as long as the dominant mixing is with the third generation \cite{BarShalom:2012ms,Geller:2012tg}.

The remainder of this paper is organized as follows.  In \Sec{sec:discoverychannels}, we compare the discovery prospects for a charged Higgs boson via top quark associated production $p p \rightarrow t H^\pm$ versus top partner decay $T \rightarrow b H^\pm$.   In \Sec{sec:detectorstudy}, we demonstrate a viable search strategy designed to uncover $T \rightarrow b H^\pm$, using realistic detector modeling and matched Monte Carlo samples to estimate the backgrounds.  We show in \Sec{sec:OtherBSM} how the same search is applicable for other scalar states that may be produced in top partners decays, such as $T \rightarrow t \varphi^0$ with $\varphi^0 \rightarrow b b$.  We conclude in \Sec{sec:Conclusion} with possible extensions of our analysis.

\section{Charged Higgs Discovery Channels}
\label{sec:discoverychannels}

Many models with extended Higgs sectors contain a charged Higgs state $H^\pm$ with a potentially large $H^\pm \rightarrow t b$ branching ratio.  For example, in a Type II two Higgs doublet model (2HDM), the absence of a measured deviation from the SM prediction for $b \rightarrow s \gamma$ indicates that  the charged Higgs bosons must be somewhat heavy, $m_{H^{\pm}} \gsim 300$ GeV \cite{Deschamps:2009rh,Chen:2013kt}, ensuring the $H^\pm \rightarrow t b$ decay mode is open.  Indeed, for such heavy charged Higgs bosons, $H^\pm \rightarrow t b$ dominates over much of the parameter space. In this paper, we assume for simplicity that the branching ratio  $\text{Br}( H^{\pm} \rightarrow tb)= 1$.  We briefly comment on the possibility of other useful decay modes in the conclusion.  We highlight the main obstacles to observing $p p \rightarrow t H^\pm$ in \Sec{sec:tHassoc}, and then discuss the potential advantages of the decay $T \rightarrow b H^\pm$ in \Sec{sec:TtobH}.  

\subsection{Via Top Quark Associated Production}
\label{sec:tHassoc}

\begin{figure}
\centering
\includegraphics[width=0.5\textwidth]{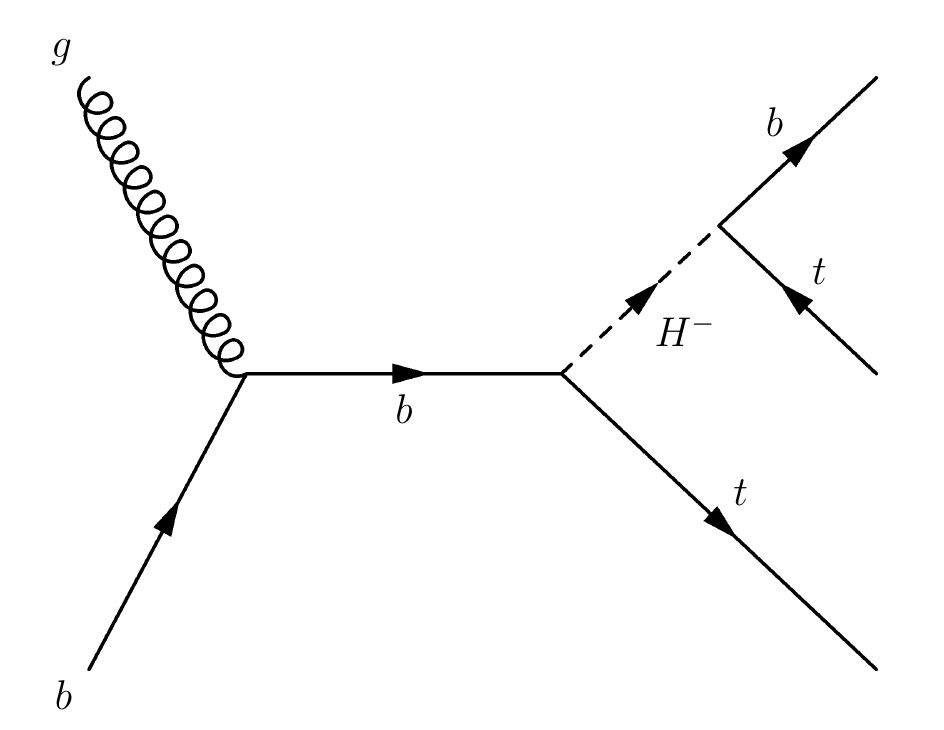}
\caption{Feynman diagram contributing to $g b \rightarrow t H^\pm$ with $H^\pm \rightarrow t b$ decay.}
\label{fig:gbtotH}
\end{figure}

There can be appreciable production of $H^\pm$ in association with a top quark via $g b \rightarrow t H^\pm$ (see \Fig{fig:gbtotH}), enabling a search for $H^\pm \rightarrow t b$ in the $t t b$ final state.  In particular, the final states in which a single top decays leptonically allow for the reconstruction of both tops (with reduced combinatoric background relative to the dileptonic or dihadronic final states) and thus the potential observation of a $H^\pm$ resonance peak in the $m_{tb}$ distribution.

Unfortunately, this channel is subject to large SM backgrounds from $tt+$jets (with a light jet faking a $b$) and $ttbb$.  One might hope that the $tt+$jets background could be avoided by requiring 3 $b$-tagged jets in the final state, as advocated in \Refs{Barger:1993th, Gunion:1993sv, Moretti:1999bw} and studied at the detector level in \Ref{Assamagan:2000uc}.  However, $tt+$jets is still a formidable background even after 3 $b$-tagged jets are required, in part because there is a relatively high charm mistag rate ($\epsilon_c \approx 0.14$ \cite{ATLAS:2011hfa, ATLAS:1999uwa} as opposed to $\epsilon_c \approx 0.01$ as assumed in \Refs{Barger:1993th, Gunion:1993sv, Moretti:1999bw,Assamagan:2000uc}), and in part because there is a non-negligible probability for QCD jet combinations to exhibit significant invariant masses (i.e.~$m_{jj} \sim m_W$ or $m_{jjj} \sim m_t$). 
Alternatively, one could attempt to search for a charged Higgs in a $ttbb$ final state from $p p \rightarrow t H^\pm b$, with the requirement of 4 $b$-tagged jets in the final state as suggested in \Ref{Miller:1999bm}.  Requiring an additional $b$-jet does suppress the $tt+$jets background.  However, the additional $b$-jet produced in $p p \rightarrow t H^\pm b$ is frequently relatively soft, suppressing the signal process if typical $b$-jet $p_{T}$ criteria are imposed.  Furthermore, even if the $tt+$jets background can be reduced to acceptable levels via this strategy, there is an irreducible background due to SM $ttbb$ production.  Consequently, even using sophisticated techniques to distinguish signal from background, the reach of this search strategy remains limited.  The discovery reach found in \Ref{Assamagan:2004tt} is $\tan \beta \gsim 50$ for $m_{H^\pm} = 500 \text{ GeV}$ in a Type II 2HDM.\footnote{\Ref{Assamagan:2004tt} assumed a conservative $b$-tagging efficiency of $\epsilon_b = 0.5$, so the reach might improve somewhat with better $b$-tagging.}  Comparing with \Ref{Dittmaier:2011ti}, this corresponds roughly to $\sigma(pp \rightarrow t H^\pm) \gsim 700 \text{ fb}$.

Thus, the discovery of a charged Higgs boson via top quark associated production seems extremely challenging, particularly for intermediate $\tan \beta$ and larger $m_{H^\pm}$.\footnote{For much larger values of $m_{H^\pm} \gsim 1 \text{ TeV}$, jet substructure techniques may offer some improvement \cite{Yang:2011jk}.}  This motivates an investigation of alternative methods for searching for charged Higgses.  

\subsection{Via Fermionic Top Partner Decays}
\label{sec:TtobH}

\begin{figure}
\centering
\includegraphics[width=\textwidth]{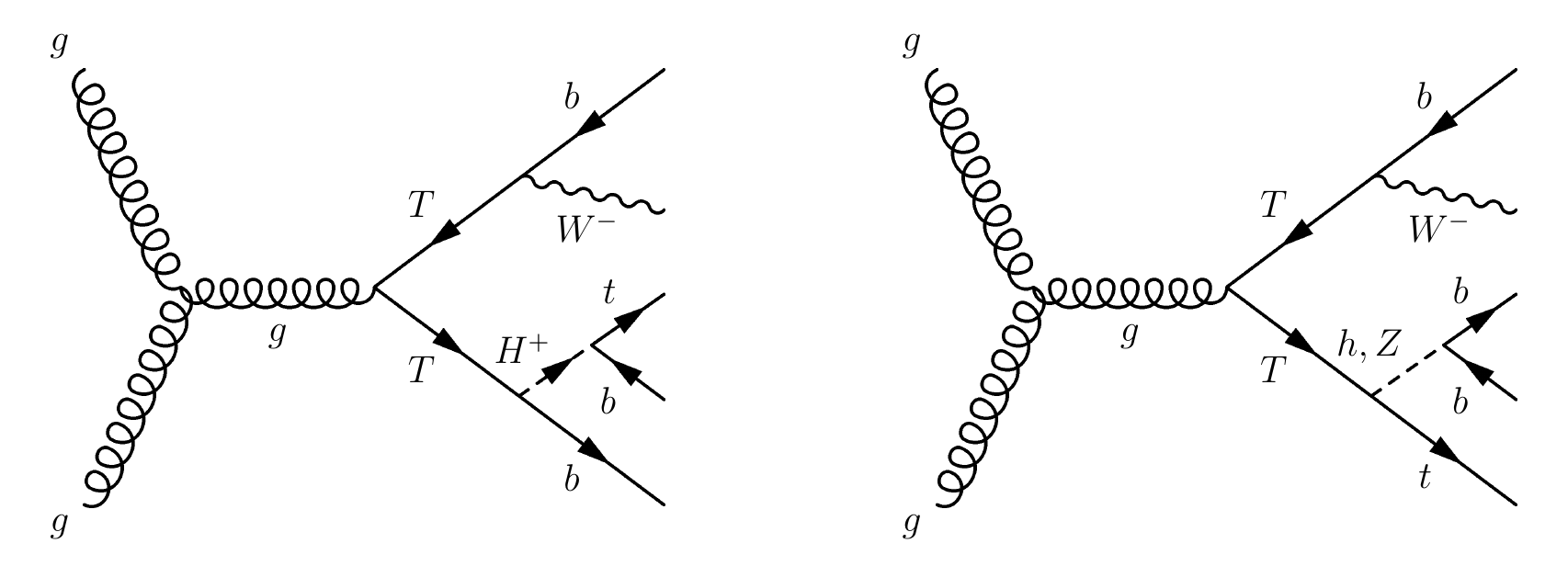}
\caption{Feynman diagrams contributing to top partner pair production, with top partners decaying to yield a $4b$, $2W^\pm$ final state.  
Our signal, containing decays of the type $T \rightarrow b H^\pm \rightarrow b t b$ (left), potentially has a background from the decays $T \rightarrow tZ, th \rightarrow tbb$ (right).}
\label{fig:TTbarpair}
\end{figure}

In this paper, we advocate an alternative method for observing $H^\pm$ at the LHC, namely in the decays of fermionic top partners.  Colored top partners can be copiously produced at hadron colliders via QCD processes $pp \to T T$ as shown in \Fig{fig:TTbarpair}.\footnote{For very large $m_T \gsim 1 \text{ TeV}$, single top partner production may dominate \cite{Han:2003wu}, favoring alternative search strategies.}  If the branching ratio for $T \rightarrow b H^\pm$ is non-negligible, top partner decays can yield a significant number of events containing at least one $H^\pm$, potentially permitting discovery.  Since the $T \rightarrow b H^\pm$ branching ratio is not necessarily suppressed at intermediate values of $\tan \beta$ (but rather depends on specific model-building details), searches in this channel can complement top quark associated production searches outlined above.  

Like the SM top fields, top partners are generally electroweak singlets or doublets, permitting renormalizable Yukawa couplings between a top partner, the Higgs field, and a SM top quark.  Consequently, top partners will typically exhibit decays to SM particles through these couplings:
\be
\label{eq:TtoSM}
T \rightarrow bW^\pm, tZ, t h.
\ee
Decays involving non-SM particles, such as $T \rightarrow b H^\pm$, are generally expected to be subdominant due to phase space suppression.  The exclusively SM decay modes in \Eq{eq:TtoSM} have been extensively studied as possible discovery channels for top partners \cite{Han:2003wu,Perelstein:2003wd,Okada:2012gy,DeSimone:2012fs}, and recent limits from the LHC have been set in \Refs{ATLAS:2012qe,Chatrchyan:2012vu}.  

We envision a scenario where the top partner $T$ is discovered---hopefully soon---via one of the decay modes in \Eq{eq:TtoSM}.  We then have the opportunity to search for subdominant decays like $T \rightarrow b H^\pm$.  In fact, when top partners are pair produced in $pp \to T T$, one can use a decay mode like $T \rightarrow bW^\pm$ to ``tag'' events as potential top partner pair events  and thereby reduce SM backgrounds (notably, events with lighter SM tops).  For concreteness, consider the event topology in \Fig{fig:TTbarpair},\footnote{For simplicity, we do not distinguish between particles and anti-particles when writing decay chains.}
\begin{equation}
\label{eq:desiredtopology}
p p \rightarrow (T \rightarrow b W_{\rm had}^\pm) (T \rightarrow b H^\pm \rightarrow b t_{\rm lep} b) \to 4b + 2j + \ell^\pm \nu,
\end{equation}
where the subscript ``had'' (``lep'') refers to decays of the corresponding $W^\pm$ to $jj$ ($\ell^\pm \nu$).  As the $W^\pm$ from the $T \rightarrow b W_{\rm had}^\pm$ decay will be relatively boosted, its hadronic decay will yield a distinctive signature of two fairly collimated jets with $m_{jj} \sim m_W$ that reconstruct a top partner with a $b$-jet.  Meanwhile, the leptonic decay on the other side of the event reduces combinatoric background, allowing a reconstruction of a second top partner in the event.

The dominant SM backgrounds are $ttbb$ and $tt+$jets with two light jets faking $b$'s.  However, the presence of four relatively hard $b$-jets in the signal means that a requirement of four $b$-tagged jets can be used (in addition to top partner reconstruction) to greatly suppress these backgrounds.  The low fake rate suppresses $tt$+jets, whereas $ttbb$ can be effectively suppressed since the additional $b$'s often come from gluon splitting, such that frequently either one $b$-jet is soft and does not pass a minimum $p_{T, j}$ requirement, or the $b$'s are collimated and consequently coalesce into a single jet.  High $b$-multiplicity requirements have similarly been applied to reduce $tt$+jets and $ttbb$ backgrounds in the context of SUSY stop searches \cite{Berenstein:2012fc} and searches for top partners decaying to exclusively SM states \cite{Harigaya:2012ir}.

With the SM background under control, a remaining challenge is that other top partner decays can yield the same final state as \Eq{eq:desiredtopology}, notably $T \rightarrow  t_{\rm{lep}}h_{bb}$ and $T \to t_{\rm{lep}}Z_{bb}$  (see \Fig{fig:TTbarpair}).   These ``background'' events exhibit a key kinematic difference, however, since the $bb$-pair from the $h$ or $Z$ is constrained to have an invariant mass of $m_{bb} = m_h$ or $m_Z$.  For signal events the $bb$ invariant mass can be much larger.  Consequently, we will see that a cut on the minimum $m_{bb}$ in the event can be used to efficiently isolate rare $T \rightarrow b H^\pm$ decays.  As long as the branching ratio $T \rightarrow b H^\pm$ is of order 10\%, then the search presented below will be sensitive to the $bH^\pm$ states.

\section{Search Strategy}
\label{sec:detectorstudy}

In this section, we describe a search strategy that can be used to discover the presence of a charged Higgs produced in $T \rightarrow bH^\pm$ based on the topology described in \Sec{eq:desiredtopology}.  As a benchmark, we choose $m_T = 700 \text{ GeV}$, a representative value that satisfies current bounds \cite{ATLAS:2012qe, Chatrchyan:2012vu,Rao:2012gf} but is not so high as to create tensions with naturalness.  Since a $H^\pm$ discovery will require high luminosity ($\simeq 300 \text{ fb}^{-1}$), we consider events for the LHC with $\sqrt{s} = 14 \text{ TeV}$.  

We first describe some of the details of our simulation framework, and then present possible event selection criteria that can identify a reasonable fraction of $T \rightarrow b H^\pm$ events while rejecting much of the SM and $T \rightarrow  th, tZ$ backgrounds.

\subsection{Simulation Framework}
\label{sec:simframework}

For our study, we use \textsc{MadGraph 5} \cite{Alwall:2011uj} to generate parton-level events, interfaced with \textsc{Pythia 6.4} \cite{Sjostrand:2006za} for decay and hadronization.   For top partner pair production, we generate MLM-matched \cite{Mangano:2006rw,Mrenna:2003if} samples of
\be
p p \rightarrow T T + nj
\ee
with $n = 0, 1, 2$ and top partners decaying as
\be
T \to b W^\pm, \quad t h,\quad t Z, \quad b H^\pm
\ee 
in \textsc{MadGraph} -- subsequent decays are carried out in \textsc{Pythia}.  Using unmatched samples, we have confirmed that we obtain similar results by (1) simulating the full $TT \rightarrow b W^\pm X \rightarrow b b b b j j \ell \nu$ ($X = b H^\pm, t h, t Z$) decay chain in \textsc{MadGraph} and (2) simulating $TT \rightarrow b W^\pm X$ in \textsc{MadGraph} with subsequent decays in \textsc{Pythia}, indicating that the latter method should indeed be sufficient for the matched samples.  For the benchmark value of $m_T = 700 \text{ GeV}$, the \textsc{MadGraph} matched cross section is
\begin{equation}
\sigma_{\rm MLM} (p p \rightarrow T T + nj, m_T = 700 \text{ GeV}) = 470 \text{ fb}.
\end{equation}
For the dominant SM backgrounds, we generate MLM-matched samples of $p p \rightarrow t t + n j$ for $n = 0, 1, 2$ in the four-flavor scheme and unmatched samples of $p p \rightarrow t t b b$.  The production cross sections from \textsc{MadGraph} for the SM processes are
\begin{align}
\label{eq:ttjjMGxs} \sigma_{\rm MLM} (p p \rightarrow t t + n j) & = 700 \text{ pb}, \\
\label{eq:ttbbMGxs} \sigma (p p \rightarrow t t b b) & = 10.3 \text{ pb}.
\end{align}

\begin{figure}
\centering
\includegraphics[width=0.7\textwidth]{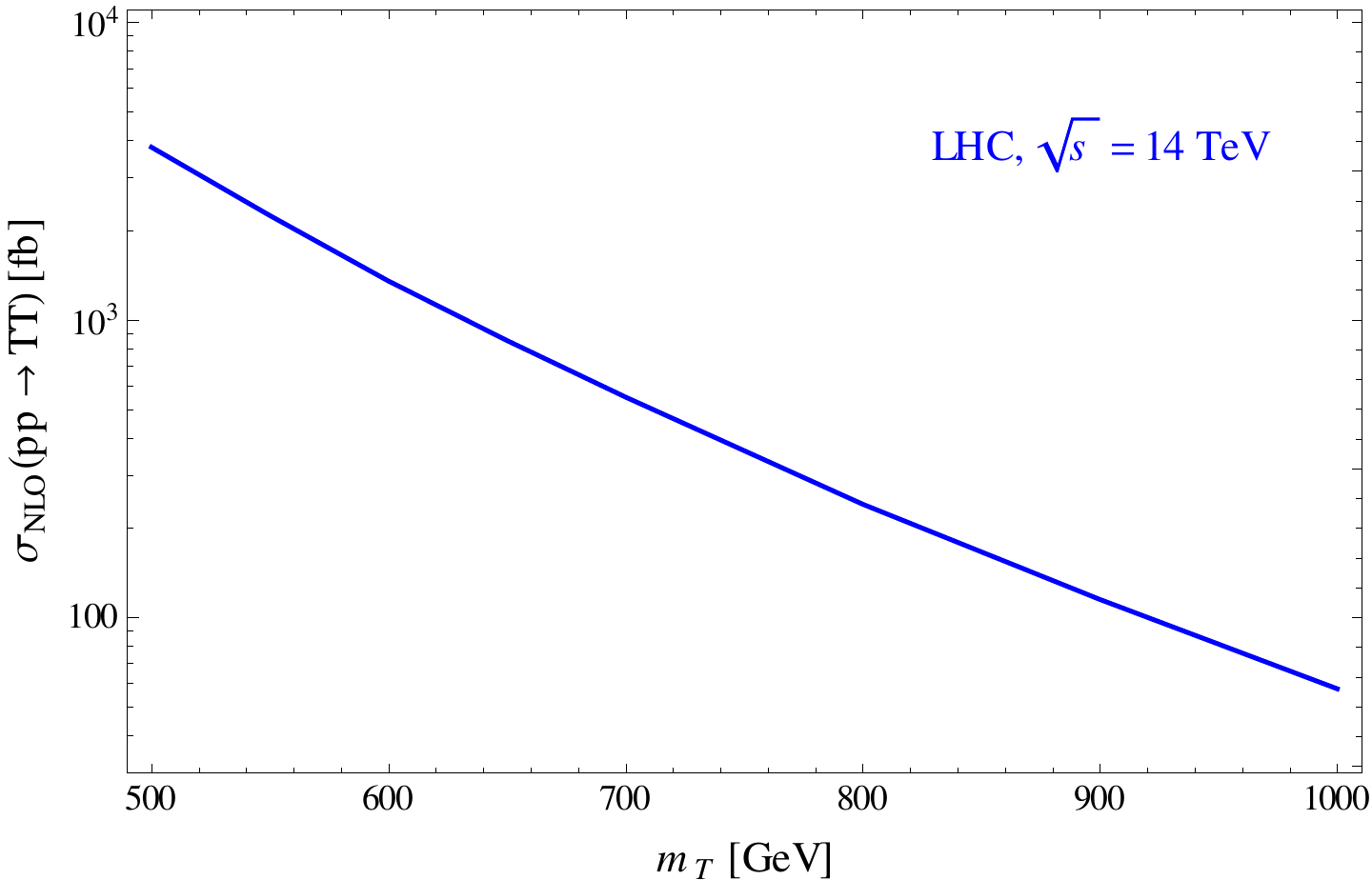}
\caption{Cross section for inclusive top partner pair production $pp \to TT$ at the LHC with $\sqrt{s} = 14 \text{ TeV}$ as a function of top partner mass $m_T$ (from \Ref{Aliev:2010zk}).  For our studies, we use the benchmark value $m_T  = 700 \text{ GeV}$.}
\label{fig:toppartnerproductionxs}
\end{figure}

All of the processes considered above are subject to sizable higher-order QCD corrections.  At NLO for the 14 TeV LHC, \textsc{Hathor} \cite{Aliev:2010zk} gives inclusive cross sections (see \Fig{fig:toppartnerproductionxs})
\begin{align}
\sigma_{\text{incl}} (p p \rightarrow t t) & = 900 \text{ pb}, \\
\sigma_{\text{incl}} (p p \rightarrow T T, m_T = 700 \text{ GeV}) & = 600 \text{ fb},
\end{align}
so we apply a $K$-factor of $K \approx 1.3$ to the $tt+$jets and $TT+$jets samples.  The appropriate $K$-factor for $ttbb$ is less readily determined, but since the $ttbb$ and $tt+$jets backgrounds are ultimately comparable, we also apply $K = 1.3$ to $ttbb$ to avoid significantly underestimating the $ttbb$ background.  As the realistic $K$-factor for $ttbb$ is likely less than that for $tt+$jets, this is a somewhat conservative choice.

Both the signal and background processes will contain two $W$ bosons from top or top partner decay.  As we will require events to contain one hard, isolated lepton which can be used to trigger the event, we allow the $W$ pair to decay via all channels capable of yielding $jj\ell \slashed{E}_T$, namely
\begin{equation}
\label{eq:Wdecays}
W W \rightarrow (jj \text{ or } \tau \nu_\tau) (\ell \nu \text{ or } \tau \nu_\tau)
\end{equation}
where the lepton or jets may arise from $\tau$ decay.  In particular, we do not account for fake leptons in this analysis, which are expected to be a small effect. 

Detector simulation was carried out using \textsc{Delphes 2.0.3} \cite{Ovyn:2009tx} (with \cite{deJeneret:2007vi, Quertenmont:2009mm}) including jet clustering with \textsc{FastJet} \cite{Cacciari:2011ma}, using resolution parameters appropriate for the ATLAS detector.  Data analysis was performed using \textsc{ROOT} \cite{Brun:1997pa}.
Electrons are required to have $p_{T, e} > 20 \text{ GeV}$ and $\abs{\eta} < 2.47$ (excluding the barrel to endcap transition region $1.37 < \abs{\eta} < 1.52$).  Muons are required to have $p_{T, \mu} > 20 \text{ GeV}$ and $\abs{\eta} < 2.5$.  Furthermore, isolation criteria are imposed.  Electrons are isolated if the transverse momentum deposited in an isolation cone of radius $\Delta R = \sqrt{(\Delta \phi)^2 + (\Delta \eta)^2} = 0.2$, $p_T^{\Delta R < 0.2} < 4 \text{ GeV}$.  Isolated muons are also required to have $p_T^{\Delta R < 0.2} < 4 \text{ GeV}$, and in addition are required to be a distance $\Delta R > 0.4$ from any jet with $p_{T, j} > 20 \text{ GeV}$ (to suppress leptons from heavy-flavor decays inside jets).  Jets are clustered using the anti-$k_T$ algorithm \cite{Cacciari:2008gp} with $R = 0.4$ and are required to have $p_{T, j} > 20 \text{ GeV}, \abs{\eta} < 2.5$.  These criteria are similar to those used in ATLAS searches for comparable final states \cite{Aad:2012qf, ATLAS:2012qe}.

For $b$-tag, light ($u, d, s$) jet mistag, and $c$-mistag efficiencies, we use the functions given in \Ref{Harigaya:2012ir} as suitable fits to the measured efficiencies \cite{ATLAS:2011hfa, ATLAS:1999uwa}, namely
\begin{align}
\epsilon_b & = 0.6 \tanh\left(\frac{p_T}{36 \text{ GeV}}\right) \times (1.02 - 0.02 \abs{\eta}), \\
\epsilon_j & = 0.001 + 0.00005 \frac{p_T}{\text{GeV}}, \\
\epsilon_c & = 0.14,
\end{align}
respectively.  In order to reduce the required number of generated events to achieve reasonable statistics (particularly for the $tt+$jets background), we consider all possible tagging configurations for any given event and weight each configuration appropriately, as opposed to implementing $b$-tagging (and mis-tagging) at the level of the detector simulation.\footnote{We do not include the effects of event pileup in this study.  Our expectation is that pileup would be most important in the reconstruction of the hadronic $W$ (see cut \ref{cut:7} below).  However, since the $W$  is at reasonably high $p_{T}$,  some additional handles, including possibly jet substructure techniques, may be able to reject fake $W$'s from pileup jets.}

\subsection{Event Selection Criteria}
\label{sec:eventselection}

The signal in \Eq{eq:desiredtopology} is characterized by a high multiplicity of relatively hard jets (including four $b$-jets), a lepton, and missing energy.  The hardest $b$ will be quite hard as it likely arises from the $T \rightarrow bW^\pm_{\rm had}$ decay.  Since the neutrino arises at the end of a longer decay chain, the signal is not characterized by particularly large missing energy, though a mild $\slashed{E}_T$ cut can still help reduce backgrounds.  We perform the following basic cuts to select events of this type:
\newcounter{numb}
\begin{enumerate}
\item \label{cut:1} Exactly 1 isolated lepton ($p_{T, \ell} > 20 \text{ GeV}$);
\item Missing energy $\slashed{E}_T > 20 \text{ GeV}$;
\item Event contains $\ge 4$ $b$-tagged jets and $\ge 2$ untagged jets ($p_{T, j} > 20 \text{ GeV}$);
\item Transverse momentum of the hardest $b$-jet satisfies $p_{T, b_1} > 160 \text{ GeV}$;
\item \label{cut:5} $m_{\text{eff}} > 1.2 \text{ TeV}$, where $m_{\text{eff}} = \sum_j p_{T,j} + p_{T,\ell} + \slashed{E}_T$, and the sum runs over all of the jets in the event.
\setcounter{numb}{\value{enumi}}
\end{enumerate}
As shown later in \Tab{tab:mH500toppartner}, these cuts reduce the SM backgrounds by orders of magnitude relative to the events containing top partners.
The exact values chosen give good top partner-to-SM background discrimination for $m_T = 700 \text{ GeV}$, but should be adjusted depending on the measured value of $m_T$ (which, as mentioned in \Sec{sec:TtobH}, we assume has been measured via a dominant decay mode).

To further suppress the $tt+$jets and $ttbb$ backgrounds and to isolate top partner pair production events containing $T \rightarrow b H^\pm$ decays, we apply the following invariant mass cuts:
\begin{enumerate}
\setcounter{enumi}{\value{numb}}
\item \label{cut:mbb}Smallest invariant mass for two $b$-tagged jets in the event satisfies $\text{min}(m_{bb}) > 150 \text{ GeV}$.  As already mentioned at the end of \Sec{sec:TtobH}, this helps suppress the background of $T \rightarrow t h$ and $T \to t Z$, but as discussed more below it also helps control the SM backgrounds.
\item \label{cut:7}Hardest $b$-tagged jet (denoted $b_1$) and two untagged jets have invariant mass $m_{b_1 j j} \approx m_T$, with the two untagged jets required to have $m_{jj} \approx m_W$ and somewhat small $\Delta R_{jj}$.  For the case of $m_T = 700 \text{ GeV}$, we require $m_{b_1 j j} \in [600,750] \text{ GeV}$ with $m_{jj} = m_W \pm 20 \text{ GeV}$ and $\Delta R_{jj} < 1.5$.
\item \label{cut:8}Event should contain three additional $b$-tagged jets (denoted $b_{2,3,4}$) that, together with the lepton and missing energy (from the neutrino), reconstruct a second top partner, i.e.~satisfying $m_{b_2 b_3 b_4 \ell \slashed{E}_T} \approx m_T$.  For $m_T = 700 \text{ GeV}$, we require $m_{b_2 b_3 b_4 \ell \slashed{E}_T} \in [500,800] \text{ GeV}$.
\end{enumerate}
The existence (or absence) of a charged Higgs state with significant coupling to top partners could be inferred from an excess (or lack of excess) of events passing these cuts.

While cut \ref{cut:mbb} was designed to reject events with $h/Z \to bb$, it is effective at rejecting $ttjj$ and $ttbb$ events as well.  For the $ttbb$ background, this is because the relatively collimated $b$'s from gluon splitting can exhibit low invariant mass.  For the $tt$+jets background, this cut rejects events where one of the quarks from the hadronic top decay is mistagged as a $b$-jet; due to the relatively large $\epsilon_c$, this can be particularly valuable in suppressing the background events with a mis-tagged charm from $W^\pm \rightarrow c s$.
In the decay of a top quark $t \rightarrow b q q^\prime$ where $q$ is mistagged as a $b$-jet 
\begin{equation}
\label{eq:mbqmistag}
m_{bq}^2 = (p_b + p_q)^2 = (p_t - p_{q^\prime})^2 = m_t^2 - 2 p_t \cdot p_{q^\prime} = m_t^2 - 2 m_t E_{q^\prime} \leq m_{t}^2,
\end{equation}
where $E_{q^\prime}$ is the energy of $q^\prime$ in the rest frame of the top quark.  So, a sufficiently hard cut on $\text{min}(m_{bb})$ can help mitigate SM backgrounds that yield the same $bbbbjj\ell \nu$ final state.  Since the majority of events are not expected to saturate the bound, we choose the cut $\text{min}(m_{bb}) > 150 \text{ GeV} > m_h, m_Z$ as a compromise between rejecting backgrounds and accepting signal events, some of which have coincidentally small $\text{min}(m_{bb})$. 

\begin{figure}[t]
\centering
\begin{subfigure}{0.49\textwidth}
\centering
\caption{\label{fig:sig}$T T \rightarrow b W_{\rm had}^\pm \, b H_{\rm lep}^\pm$} 
\includegraphics[width=\textwidth]{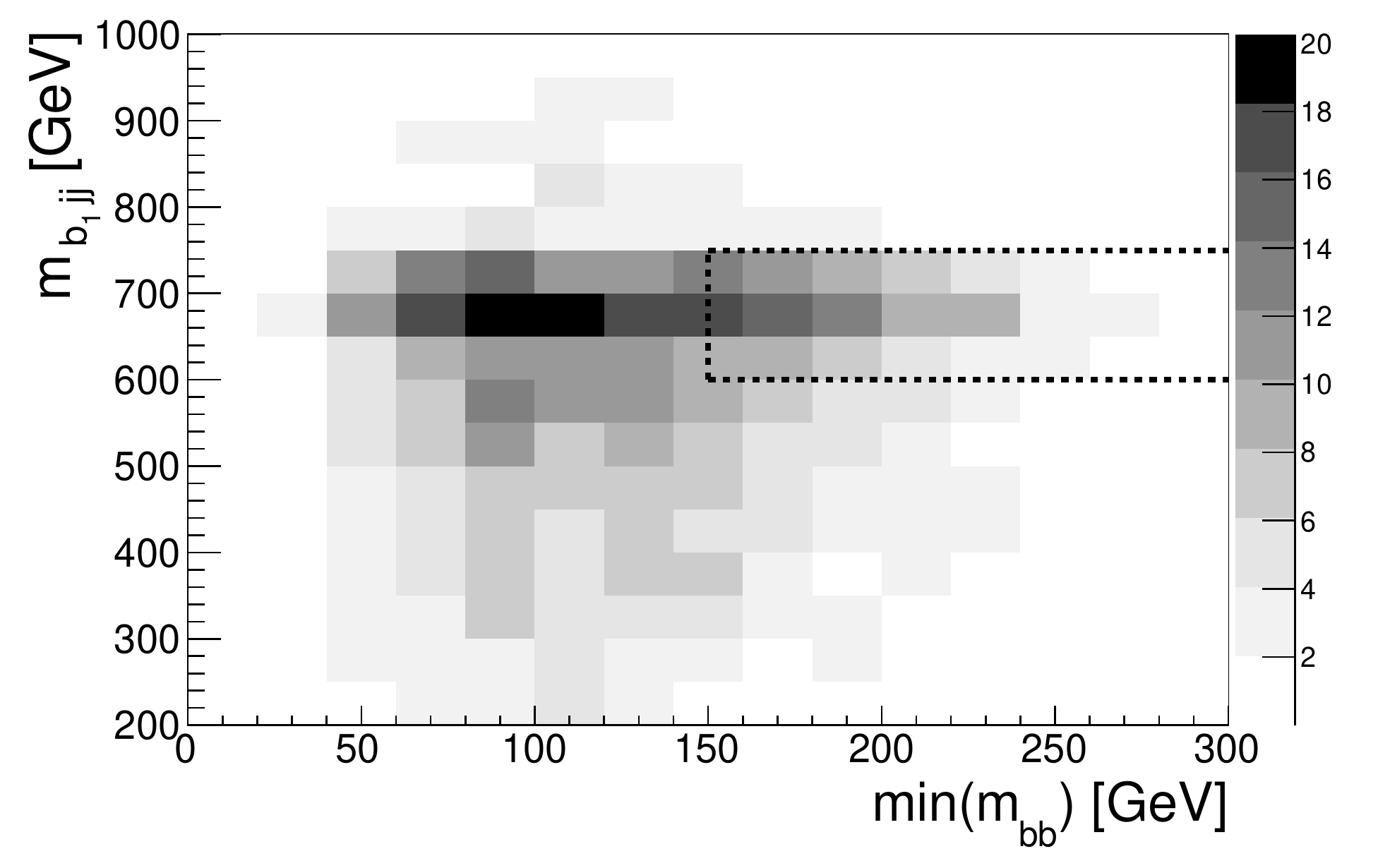}
\end{subfigure}
\begin{subfigure}{0.49\textwidth}
\centering
\caption{\label{fig:thbckgd}$T T \rightarrow b W^\pm_{\rm had} \, t_{\rm lep} h_{bb}$}
\includegraphics[width=\textwidth]{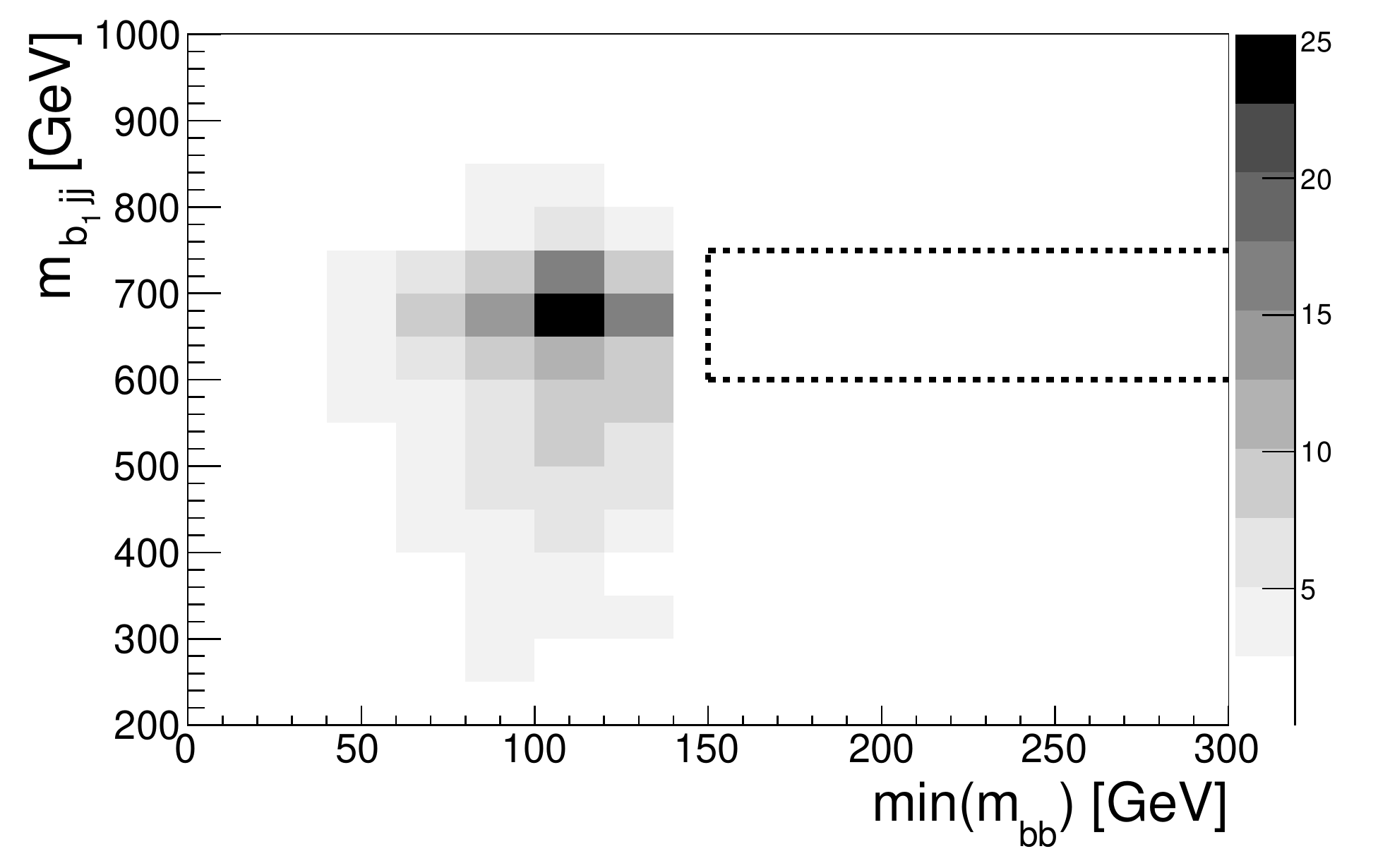}
\end{subfigure}
\vspace{.15in}
\begin{subfigure}{0.49\textwidth}
\centering
\caption{\label{fig:tZbckgd}$T T \rightarrow b W^\pm_{\rm had} \, t_{\rm lep} Z_{bb}$} 
\includegraphics[width=\textwidth]{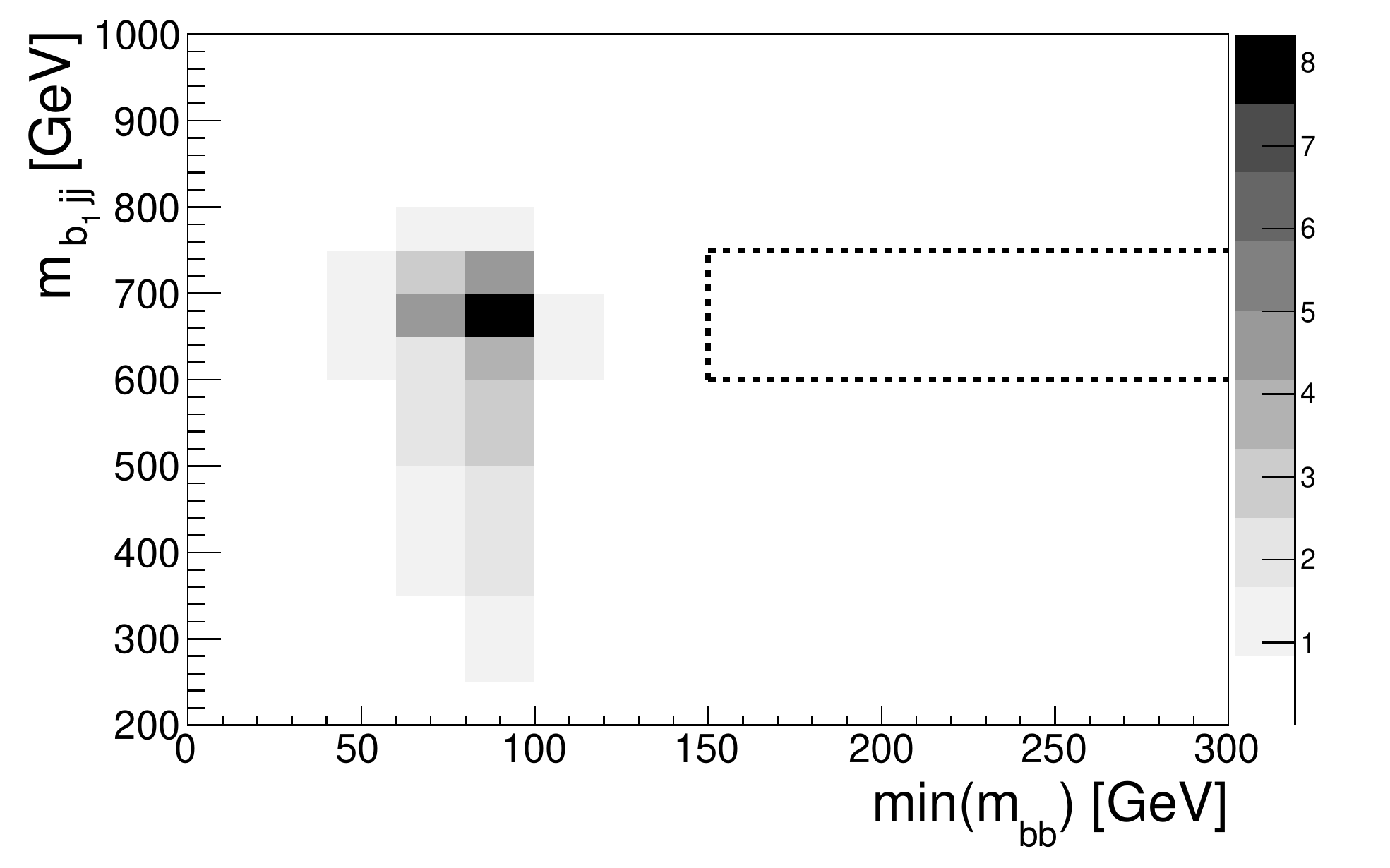}
\end{subfigure}
\begin{subfigure}{0.49\textwidth}
\centering
\caption{\label{fig:lepthbckgd}$T T \rightarrow b W^\pm_{\rm lep} t_{\rm had} h_{bb}$}
\includegraphics[width=\textwidth]{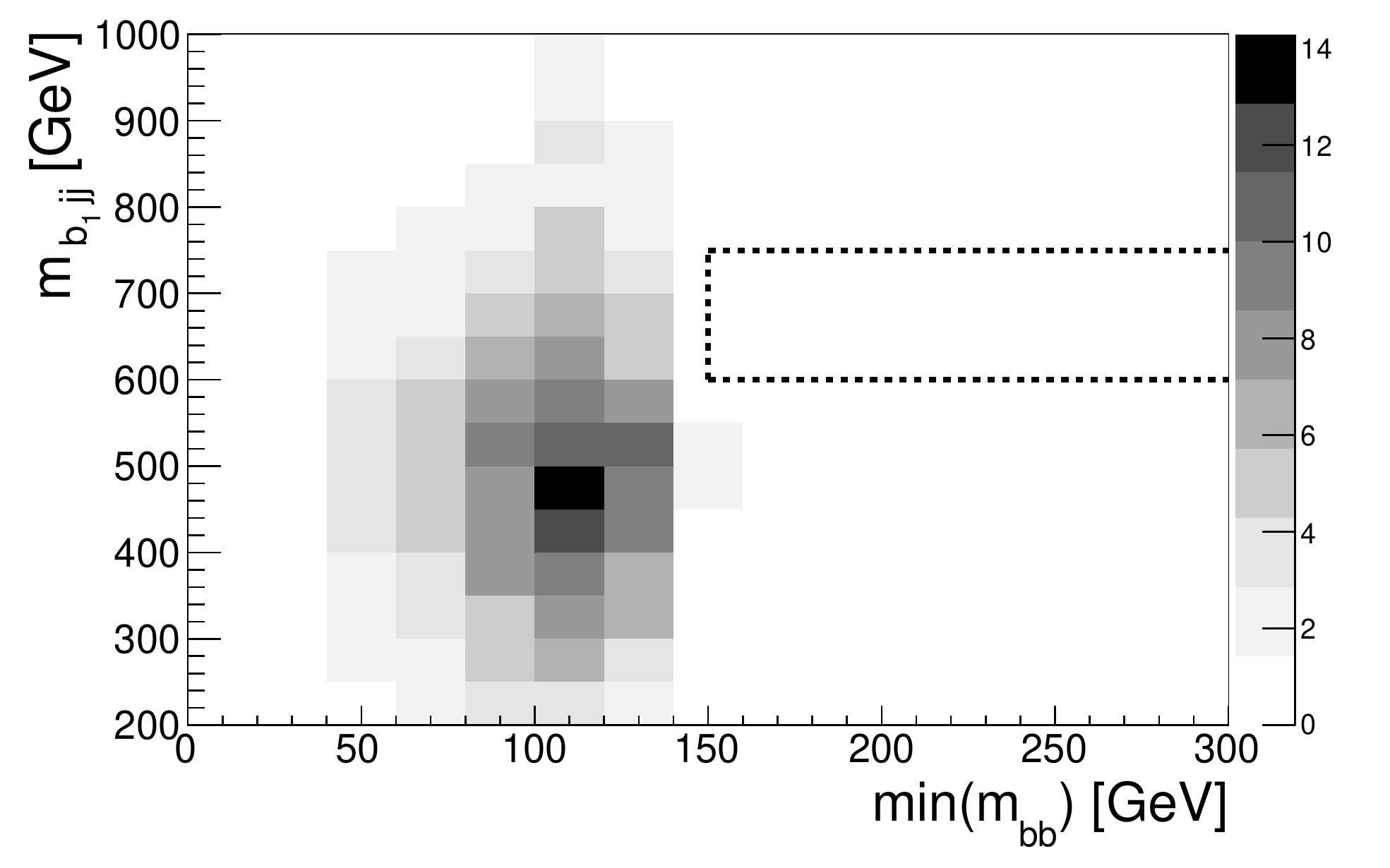}
\end{subfigure}
\vspace{.15in}
\begin{subfigure}{0.49\textwidth}
\centering
\caption{\label{fig:ttjjbckgd}$tt+$jets}
\includegraphics[width=\textwidth]{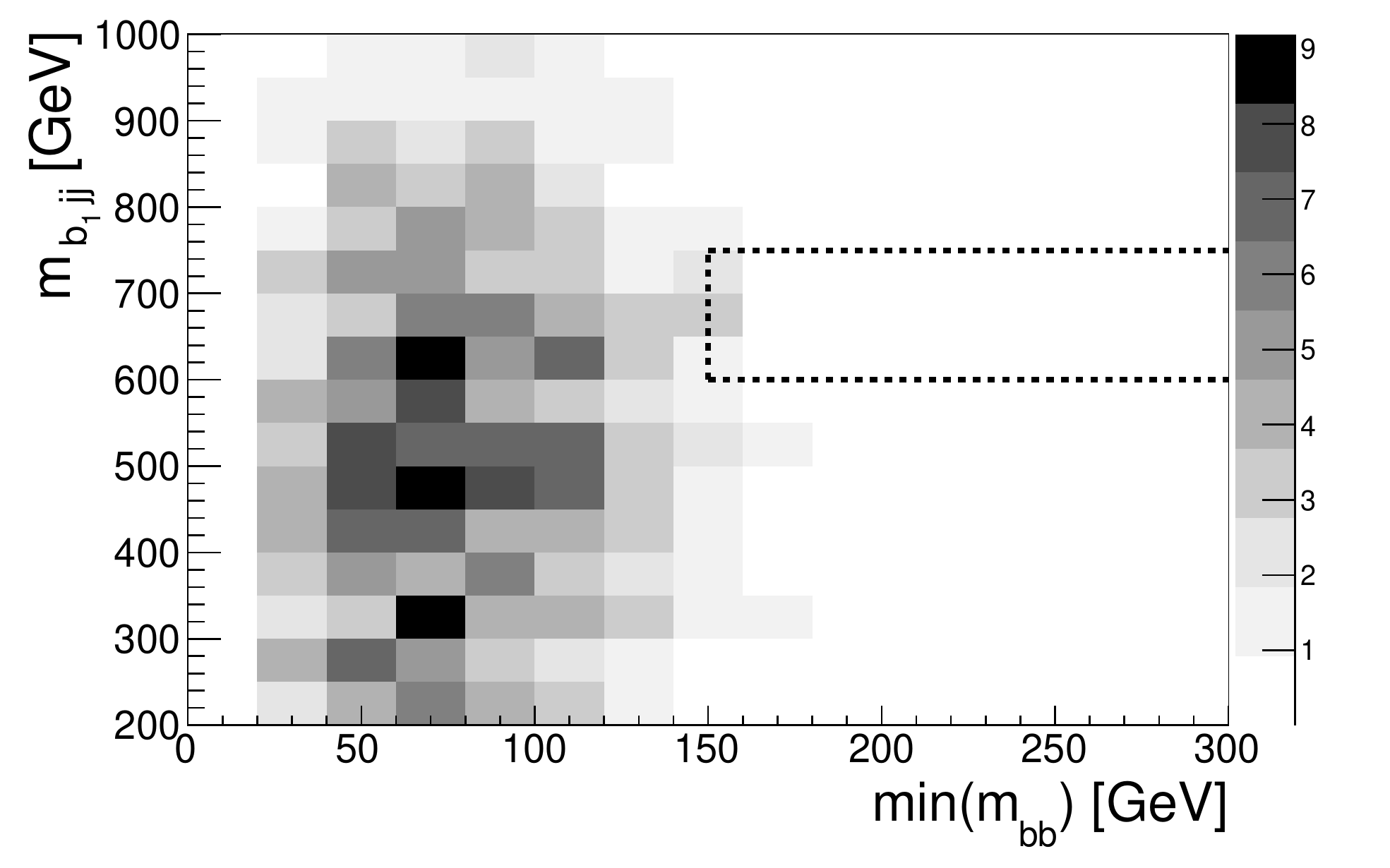}
\end{subfigure}
\begin{subfigure}{0.49\textwidth}
\centering
\caption{\label{fig:ttbbbckgd}$ttbb$}
\includegraphics[width=\textwidth]{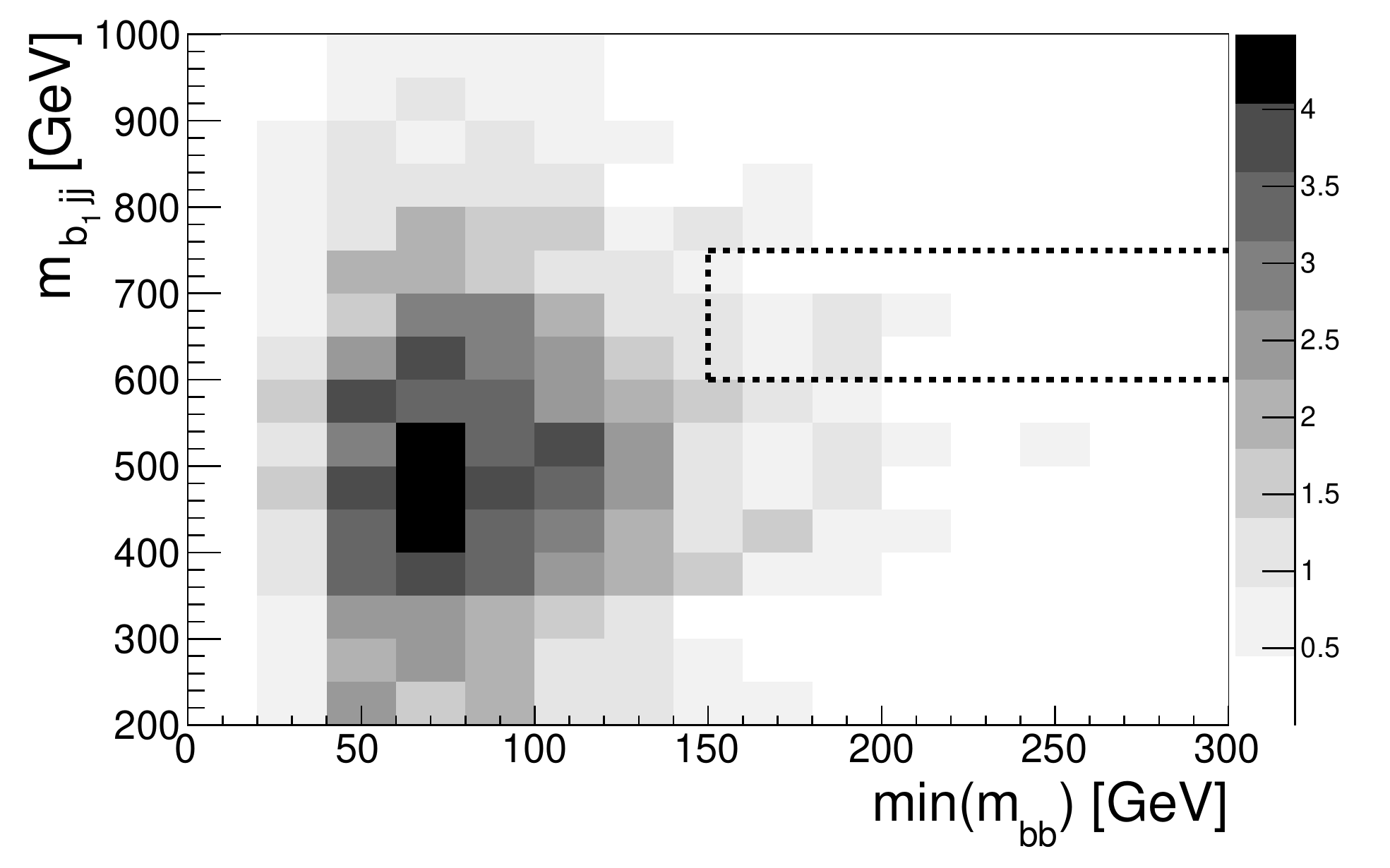}
\end{subfigure}

\caption{Distributions of $\text{min}(m_{bb})$ against $m_{b_1 jj}$ after applying basic cuts (\ref{cut:1}--\ref{cut:5}), for $m_T = 700 \text{ GeV}, m_{H^\pm} = 500 \text{ GeV}$.  Here, $m_{b_1 jj}$ corresponds to all untagged jet pairs satisfying $m_{jj} = m_W \pm 20 \text{ GeV}$ and $\Delta R_{jj} < 1.5$.  Dashed lines denote the signal region (cuts \ref{cut:mbb} and \ref{cut:7}).   For \Fig{fig:sig} through \Fig{fig:lepthbckgd}, grayscale represent Events/$\text{Br}_{bW X}$ $[300 \text{ fb}^{-1}]$, where $\text{Br}_{bW X}$ denotes the branching ratio for the process $TT \rightarrow bW^\pm X$.  For \Fig{fig:ttjjbckgd} and \Fig{fig:ttbbbckgd}, grayscale represents Events $[300 \text{ fb}^{-1}]$.}
\label{fig:2dmbbmbjj}
\end{figure}

To demonstrate how these invariant mass cuts are effective, \Fig{fig:2dmbbmbjj} shows distributions of $\text{min}(m_{bb})$ (cut \ref{cut:mbb}) versus $m_{b_1 jj}$ (cut \ref{cut:7})  for a variety of top partner processes and SM backgrounds after applying only basic cuts.\footnote{As described in cut \ref{cut:7}, $m_{b_1 jj}$ is only shown if there is an untagged jet pair satisfying $m_{jj} = m_W \pm 20 \text{ GeV}$ and $\Delta R_{jj} < 1.5$.}  We take $m_T = 700 \text{ GeV}$ and $m_{H^\pm} = 500 \text{ GeV}$, and the benchmark cuts maintain a good fraction of the signal topology in \Eq{eq:desiredtopology}.  The cut on $m_{b_1 jj} \approx m_T$ serves to isolate top partner events with a $T \rightarrow b W_{\rm had}^\pm$ decay.  The top partner clearly shows up as a band in the $m_{b_{1} jj}$ distribution in panels \Fig{fig:sig}--\Fig{fig:tZbckgd}.  Furthermore, whereas \Fig{fig:thbckgd} and \Fig{fig:tZbckgd} are peaked at $(m_{b_1jj}, \text{min}(m_{bb})) \approx (m_T, m_{h,Z})$, \Fig{fig:sig} exhibits a band at $m_{b_1jj} \approx m_T$ with $\text{min}(m_{bb})$ extending over a range of values including $\text{min}(m_{bb}) > m_{h,Z}$.  As a result, the cut on $\text{min}(m_{bb})$ isolates the $T \rightarrow b H^\pm$ decay from other top partner decays.  Also \Fig{fig:ttjjbckgd} and \Fig{fig:ttbbbckgd} demonstrate the efficacy of the $m_{bb}$ cut against the SM backgrounds for the reasons described above.
The process
\begin{equation}
\label{eq:wrongdecays}
p p  \rightarrow (T \rightarrow b W_{\rm lep}^\pm) (T \rightarrow b H_{\rm had}^\pm)
\end{equation}
is largely rejected by our cuts, but is counted as signal as it involves a charged Higgs.\footnote{In principle, one could enhance the signal sensitivity by crafting a selection criteria designed for \Eq{eq:wrongdecays}.  We found only a marginal improvement, however, since it is harder to develop a good $T \rightarrow b W_{\rm lep}^\pm$ tag to reject the $tt+$jets background.}

\begin{figure}
\centering
\includegraphics[width=0.7\textwidth]{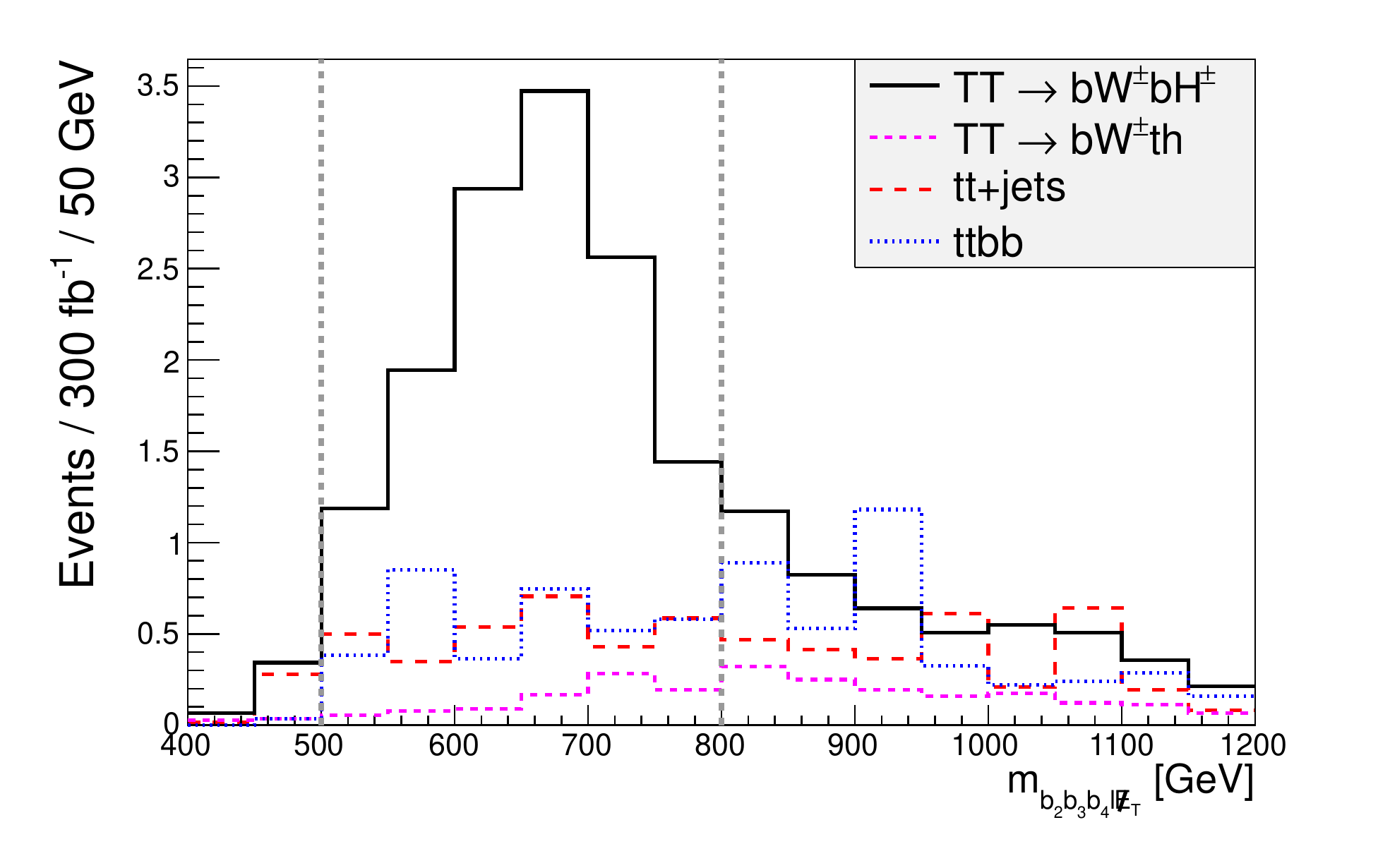}
\caption{Distribution of $m_{b_2 b_3 b_4 \ell \slashed{E}_T}$ after cuts \ref{cut:1} through \ref{cut:7} have been applied.  As in \Fig{fig:2dmbbmbjj}, $m_T = 700 \text{ GeV}$ and $m_{H^\pm} = 500 \text{ GeV}$, and in addition we take $\text{Br}_{bW bH^\pm} = 0.1$ and $\text{Br}_{bW th} = 0.2$.  The shape of the distribution for $TT \rightarrow bW^\pm tZ$ is similar to that for $TT \rightarrow bW^\pm th$.  Dashed lines denote the region selected by cut \ref{cut:8}, $m_{b_2 b_3 b_4 \ell \slashed{E}_T} \in [500,800] \text{ GeV}$.}
\label{fig:cut8plot}
\end{figure}

Distributions of $m_{b_2 b_3 b_4 \ell \slashed{E}_T}$ are shown in \Fig{fig:cut8plot} for the signal $TT \rightarrow bW^\pm bH^\pm$ and dominant SM background processes after cuts \ref{cut:1} through \ref{cut:7} have been applied.  The presence of a resonance structure at $m_{b_2 b_3 b_4 \ell \slashed{E}_T} \approx m_T$ in the signal distribution means that cut \ref{cut:8} on $m_{b_2 b_3 b_4 \ell \slashed{E}_T}$ can be used to isolate events with a second top partner and further reduce the SM backgrounds.  Note that the sharpness of the signal peak is enhanced by cut \ref{cut:7} which helps to resolve combinatoric ambiguity.

\subsection{Results}

\begin{table}
\centering
\begin{tabular}{c l | c | *2{c} | *2{c}}
\multicolumn{2}{c |}{\multirow{2}{*}{Process}} & $T T \rightarrow$ & \multicolumn{2}{c |}{$T T \rightarrow$} & \multicolumn{2}{c}{SM} \\ & & $b W^\pm b H^\pm$ & $b W^\pm t h$ & $b W^\pm t Z$ & $t t + nj$ & $t t b b$ \\ \hline
\multicolumn{2}{c |}{$\sigma \times \text{Br}$ [fb]} & 300 $\text{Br}_{bW bH^\pm}$ & 170 $\text{Br}_{bW th}$ & 44 $\text{Br}_{bW tZ}$ & $4.4 \times 10^5$ & $6.6 \times 10^3$ \\ \hline
\multicolumn{2}{c |}{Basic Cuts} & $3.6 \times 10^{-2}$ & $3.0 \times 10^{-2}$ & $2.6 \times 10^{-2}$ & $7.4 \times 10^{-6}$ & $1.6 \times 10^{-4}$ \\
Cut \ref{cut:mbb}: & $\text{min}(m_{bb})$
 & $1.3 \times 10^{-2}$ & $2.4 \times 10^{-3}$ & $2.1 \times 10^{-3}$ & $6.1 \times 10^{-7}$ & $2.8 \times 10^{-5}$ \\
Cut \ref{cut:7}: & $m_{b_1 j j}$ 
& $2.2 \times 10^{-3}$ & $2.6 \times 10^{-4}$ & $2.3 \times 10^{-4}$ & $5.4 \times 10^{-8}$ & $4.1 \times 10^{-6}$ \\
Cut \ref{cut:8}: & $m_{b_2 b_3 b_4 \ell \slashed{E}_T}$
& $1.5 \times 10^{-3}$ & $8.4 \times 10^{-5}$ & $5.7 \times 10^{-5}$ & $2.3 \times 10^{-8}$ & $1.7 \times 10^{-6}$ \\ \hline
\multicolumn{2}{c |}{Events [$300 \text{ fb}^{-1}$]} & 130 $\text{Br}_{bW bH^\pm}$ & 4.3 $\text{Br}_{bW th}$ & 0.76 $\text{Br}_{bW tZ}$ & 3.1 & 3.4
\end{tabular}
\caption{\label{tab:mH500toppartner} Cumulative efficiencies for signal and background events to pass the selection criteria.  Signals are generated for a representative heavy charged Higgs mass, $m_{H^\pm} = 500 \text{ GeV}$.  In all events, $W^\pm$ bosons decay as specified in \Eq{eq:Wdecays}, and the Higgs and $Z$ bosons in these events decay to $bb$.  We take $\text{Br}(h \rightarrow bb) = 0.58$, $\text{Br}(Z \rightarrow bb) = 0.15$, and assume $\text{Br}(H^\pm \rightarrow t b) = 1$.  $\text{Br}_{bW X}$ denotes the branching ratio for $TT \rightarrow bW^\pm X$.  
The cut ranges are defined as $\text{min}(m_{bb}) > 150 \text{ GeV}$ (cut \ref{cut:mbb}), $m_{b_1 j j} \in [600,750] \text{ GeV}$ (cut \ref{cut:7}), and $m_{b_2 b_3 b_4 \ell \slashed{E}_T} \in [500,800]$ (cut \ref{cut:8}).}
\end{table}

Efficiencies for the various cuts from \Sec{sec:eventselection} are shown in \Tab{tab:mH500toppartner} for a representative heavy charged Higgs mass, $m_{H^\pm} = 500 \text{ GeV}$.  For these efficiencies, the SM background contributions from $tt+$jets and $ttbb$ are comparable.
Also shown are the dominant background contributions arising from decays of top partners to electroweak bosons.  In principle, top quark associated production of $H^\pm$ is also a ``background" (as it does not serve our goal of uncovering information about the $H^\pm$ coupling to top partners), but it tends to be negligible unless $\sigma(p p \rightarrow t H^\pm) \gsim \mathcal{O}(600) \text{ fb}$.  In terms of the complementarity of these two channels as methods for searching for $H^\pm$, it is worth noting that this is exactly the region in which a top quark associated production search becomes potentially viable, see \Sec{sec:tHassoc}.

The discovery potential of this search depends on the branching ratios of the top partners.  As an illustrative example, consider the parametrization
\be
\label{eq:epsilondef}
T \to \left\{\begin{array}{ll}b H^\pm & \quad \text{Br} = \epsilon \\ bW^\pm & \quad \text{Br} = \frac{1}{2}(1-\epsilon) \\t Z & \quad \text{Br} = \frac{1}{4}(1-\epsilon)\\ t h & \quad \text{Br} = \frac{1}{4}(1-\epsilon) \end{array}\right. .
\ee
The  $2:1:1$ ratio for the $bW^\pm:tZ:th$ modes is what one might approximately expect due to the Goldstone Boson Equivalence Theorem \cite{Cornwall:1974km,Vayonakis:1976vz,Lee:1977eg}.  Using the efficiencies in \Tab{tab:mH500toppartner} for $m_T = 700 \text{ GeV}, m_{H^\pm} = 500 \text{ GeV}$, we find using Poisson statistics that with $\mathcal{L} = 300 \text{ fb}^{-1}$ of integrated data, one can probe
\begin{equation}
\epsilon = \left\{\begin{array}{c c c l} 0.04 & \text{ at } & 2 \sigma & (S = 5.5, B = 6.5_{\text{SM}} + 1.2 = 7.7) \\ 0.12 & \text{ at } & 5 \sigma & (S = 13.7, B = 6.5_{\text{SM}} + 1.0 = 7.5) \end{array}\right.,
\end{equation}
indicating that this channel is viable even for relatively modest $T \rightarrow b H^\pm$ branching ratios.  The change in $B$ results from the change in $\text{Br}_{bW th, bW tZ}$ as a function of  $\epsilon$, i.e. these decay processes contribute an expected 1.2 background events at $\epsilon = 0.04$ but 1.0 events at an $\epsilon=0.12$.   In realistic 2HDMs with fermionic top partners, such as the ``Bestest Little Higgs'' \cite{Schmaltz:2010ac}, a wide variety of decay branching ratios are possible for the various top partners in different regions of parameter space, making this channel worthy of exploration if fermionic top partners are discovered (for a sense of the various branching ratios possible in the ``Bestest Little Higgs,'' see \Ref{Godfrey:2012tf}).  

As we consider a signal process involving $T \rightarrow b W^\pm \rightarrow b j j$, there is also in principle an upper limit on the $\epsilon$ that can be probed using this approach, above which the channel would be suppressed by small $\text{Br}(T \rightarrow b W^\pm)$.  We view this possibility as unlikely because, as mentioned, the $T \rightarrow b H^\pm$ decay is likely to be subdominant due to phase space suppression.  If the $T \rightarrow b H^\pm$ decay does dominate, alternative search strategies would likely be preferred to tease out the existence of the $H^{\pm}$. However,  such top partners would at least be discovered via the kinds of multi-$b$ searches used to hunt for $T \rightarrow th$ final states, as long as no $m_{bb} = m_{h}$ requirement is  applied.

\begin{table}
\centering
\begin{tabular}{c c | c c c c}
$m_T$ & $m_{H^\pm}$ & Efficiency & Events [$\mathcal{L} = 300~\text{fb}^{-1}$] & $\epsilon \; (2 \sigma)$ & $\epsilon \; (5 \sigma)$ \\ \hline
\multirow{3}{*}{700} & 400 & $1.5 \times 10^{-3}$ & 130 $\text{Br}_{bW bH^\pm}$ & 0.04 & 0.12  \\
& 500 & $1.5 \times 10^{-3}$ & 130 $\text{Br}_{bW bH^\pm}$ & 0.04 & 0.12 \\
& 600 & $8.2 \times 10^{-4}$ & 73 $\text{Br}_{bW bH^\pm}$ & 0.08 & 0.24
\end{tabular}
\caption{\label{tab:toppartnerdifferentmasses} Efficiencies for passing the given selection criteria for $m_T = 700 \text{ GeV}$ and several representative values of $m_{H^\pm}$.  Also shown are corresponding values of $\epsilon$ (defined in \Eq{eq:epsilondef}) yielding $2 \sigma$ and $5 \sigma$ significance assuming $\text{Br}(H^\pm \rightarrow t b) = 1$ and $\mathcal{L} = 300~\text{fb}^{-1}$.  The $2 \sigma$ ($5 \sigma$) significances correspond to $S \approx 5.5 \; (13.7)$ and $B \approx 7.7 \; (7.5)$.}
\end{table}

Efficiencies for passing the given selection criteria, and corresponding values of $\epsilon$ yielding $2 \sigma$ and $5 \sigma$ significances with the branching ratios described above, are given in \Tab{tab:toppartnerdifferentmasses} for several representative values of $m_{H^\pm}$.  
For $m_{H^\pm} \approx m_T$, the efficiency for the signal process to pass the selection criteria falls because the $b$ quark from $T \rightarrow b H^{\pm}$ becomes softer, increasing the likelihood of an event failing cut \ref{cut:mbb} by having $\text{min}(m_{bb}) < 150 \text{ GeV}$.  Thus, in these regions of parameter space, a larger $T \rightarrow b H^{\pm}$ branching ratio is required for this to be a viable search strategy -- unfortunately, also in these regions, the phase space suppression of $T \rightarrow b H^{\pm}$ will be greater, likely reducing this branching ratio.  For optimal coverage of this squeezed region, it might be worth pursuing a set of  dedicated cuts.  For larger values of $m_T$, we anticipate that comparable separation from SM backgrounds could be achieved with slightly looser cuts due to the increased hardness of the event.  The corresponding increase in efficiency could partially mitigate the rapid decrease in $\sigma_{\rm NLO}(pp \rightarrow TT)$ with $m_T$ (\Fig{fig:toppartnerproductionxs}).

\begin{table}
\centering
\begin{tabular}{c c | c c c c}
$m_T$ & $m_{H^\pm}$ & Efficiency & Events [$\mathcal{L} = 3000~\text{fb}^{-1}$] & $\epsilon \; (2 \sigma)$ & $\epsilon \; (5 \sigma)$ \\ \hline
\multirow{3}{*}{1000} & 400 & $1.2 \times 10^{-3}$ & 110 $\text{Br}_{bW bH^\pm}$ & 0.07 & 0.19  \\
& 600 & $1.7 \times 10^{-3}$ & 150 $\text{Br}_{bW bH^\pm}$ & 0.05 & 0.13 \\
& 800 & $1.4 \times 10^{-3}$ & 120 $\text{Br}_{bW bH^\pm}$ & 0.06 & 0.17
\end{tabular}
\caption{\label{tab:toppartnerdifferentmassesMT1000} Efficiencies for passing the given selection criteria for $m_T = 1 \text{ TeV}$ and several representative values of $m_{H^\pm}$.  Also shown are corresponding values of $\epsilon$ (defined in \Eq{eq:epsilondef}) yielding $2 \sigma$ and $5 \sigma$ significance assuming $\text{Br}(H^\pm \rightarrow t b) = 1$ and $\mathcal{L} = 3000~\text{fb}^{-1}$.  In this case, we require $m_{b_1 j j} \in [900,1050] \text{ GeV}$ and $m_{b_2 b_3 b_4 \ell \slashed{E}_T} \in [800,1100] \text{ GeV}$.  For these cuts, the $tt$+jets and $ttbb$ SM processes contribute 6.9 and 3.9 background events, respectively.  The $2 \sigma$ ($5 \sigma$) significances correspond to $S \approx 7.0 \; (17.2)$ and $B \approx 11.8 \; (11.6)$.}
\end{table}

To demonstrate the potential reach of this search at the LHC with very high luminosity, we present the analog of \Tab{tab:toppartnerdifferentmasses} for $m_T = 1 \text{ TeV}$ and $\mathcal{L} = 3000 \text{ fb}^{-1}$ in \Tab{tab:toppartnerdifferentmassesMT1000}.  The increase in luminosity is necessary to compensate for the decrease in production cross section,
\begin{align}
\sigma_{\text{incl}} (p p \rightarrow T T, m_T = 1 \text{ TeV}) & = 60 \text{ fb}.
\end{align}
In this case, we modify cuts \ref{cut:7} and \ref{cut:8} to require $m_{b_1 j j} \in [900,1050] \text{ GeV}$ and $m_{b_2 b_3 b_4 \ell \slashed{E}_T} \in [800,1100] \text{ GeV}$.  Ideally, however, the other cuts would also be optimized for $m_T = 1 \text{ TeV}$.  For instance, heavier top partners produce events with larger $p_{T,b_1}$ and $m_{\text{eff}}$, such that harsher basic cuts may be preferred to further suppress SM backgrounds.  As the $W^\pm$ from the $T \rightarrow b W^\pm_{\text{had}}$ decay would be more boosted, cut \ref{cut:7} could also be modified to require more collimated jets---jet substructure techniques may even prove useful in this regime.  Finally, as heavier top partners permit more phase space for decays, the $\text{min}(m_{bb})$ required could conceivably be increased.  Appropriately optimizing cuts for different candidate values of $m_T$ would extend the reach of this search.

\begin{figure}
\centering
\includegraphics[width=0.7\textwidth]{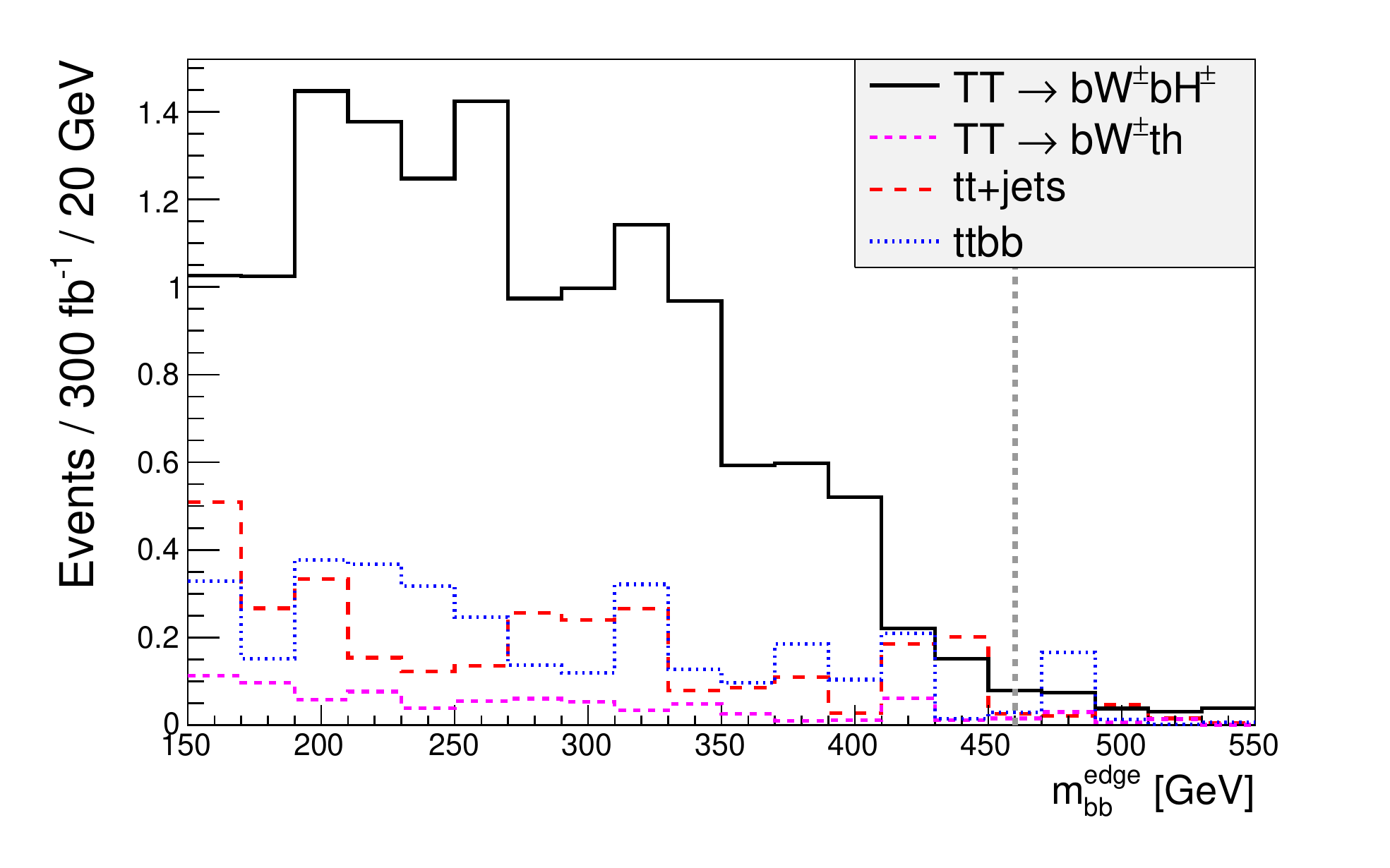}
\caption{Distribution of $m_{bb}^{\text{edge}}$ taking $m_T = 700 \text{ GeV}, m_{H^\pm} = 500  \text{ GeV}, \epsilon = 0.12$.  The $TT \rightarrow bW^\pm tZ$ distribution is not shown as it is similar in shape to the $TT \rightarrow bW^\pm th$ distribution, but is suppressed as $\text{Br}(Z \rightarrow bb) < \text{Br}(h \rightarrow bb)$.  For these values, the $b$'s from $T \rightarrow H^\pm b \rightarrow t b b$ are constrained to have $m_{bb}^{\rm{edge}} \leq 460 \text{ GeV}$ (dashed line, see \Eq{eq:mbbbound}).} 
\label{fig:mbbedgeplot}\end{figure}

The above analysis strategy was aimed at getting a signal to background ratio of $\mathcal{O}(1)$, so relatively harsh cuts were needed to control the SM background from top quarks.  One drawback of this analysis strategy is that the number of signal events passing these criteria is likely to be small, precluding the observation of, e.g., a resonance peak at $m_{tb} = m_{H^\pm}$.  Multivariate techniques may extend the discovery potential of this search, but are unlikely to increase event yields sufficiently to allow for the determination of $m_{H^\pm}$ unless looser event selection criteria (and alternative ways of controlling the SM top backgrounds) are used.  However, with sufficient data, there are numerous methods through which the charged Higgs mass could be extracted from this channel, even if $H^\pm$ has leptonic decays.  For example, one way to access the $H^{\pm}$ mass is via the edge in the $m_{bb}$ distribution for the $b$'s produced in the decay $T \rightarrow H^\pm b \rightarrow tbb$,  
\begin{equation}
\label{eq:mbbbound}
m_{bb} \leq m_T \sqrt{1 - \frac{m_{H^\pm}^2}{m_T^2}} \sqrt{1 - \frac{m_t^2}{m_{H^\pm}^2}}.
\end{equation}
This, too is likely to be challenging due to small statistics, but given lighter top partners, a sufficiently large data set, or generous branching ratios, it could be worth pursuing further. To give an idea of how this might work, we first attempt to identify the $b$ quark coming from the top decay by minimizing $|m_{b_k \ell \slashed{E}_T} -m_{t}|$ ($k=2,3,4$, i.e.~excluding the harder $b$ used in the other side $T$ reconstruction).  We denote this $b$ as $b_{t}$.  We can then examine the invariant mass distribution of the remaining two $b$ quarks: $m_{bb}^{\rm{edge}}$.  A sample distribution is shown for $m_T = 700 \text{ GeV}$, $m_{H^\pm} = 500 \text{ GeV}$, and $\epsilon = 0.12$ in \Fig{fig:mbbedgeplot}.  For these values, $m_{bb}^{\rm{edge}} \leq 460 \text{ GeV}$.  Unlike attempting to observe a resonance in an $m_{tb}$ distribution, the $m_{bb}^{\rm{edge}}$ distribution has the advantage of not being subject to combinatoric ambiguity once $m_{b_t \ell \slashed{E}_T} \approx m_t$ has been used to identify the bottom arising from the leptonic top quark decay.

\section{Applicability to Neutral Scalars}
\label{sec:OtherBSM}

The strategy outlined above is clearly suitable for searching for any charged scalars $\varphi^\pm$ produced in top partner decays $T \rightarrow b \varphi^\pm$ with $\varphi^\pm \rightarrow t b$.  However, it is also applicable to heavier neutral scalar states $\varphi^0$ produced via $T \rightarrow t \varphi^0$ and decaying as $\varphi^0 \rightarrow b b$,
\begin{equation}
\label{eq:alttopology}
p p \rightarrow (T \rightarrow b W_{\rm had}^\pm) (T \rightarrow t \varphi^0 \rightarrow t_{\rm lep} b b) \to 4b + 2j + \ell \nu.
\end{equation}
While one could imagine other dedicated searches for such a $\varphi^0$, the search strategy provided already for $H^\pm$ would at least uncover an excess as long as $m_{\varphi^0} > 150 \text{ GeV}$ to satisfy the conditions of cut \ref{cut:mbb}.

\begin{table}
\centering
\begin{tabular}{c c | c c c c}
$m_T$ & $m_{\varphi^0}$ & Efficiency & Events [$300 \text{ fb}^{-1}$] & $\epsilon \; (2 \sigma)$ & $\epsilon \; (5 \sigma)$ \\ \hline
\multirow{2}{*}{700} & 350 & $1.3 \times 10^{-3}$ & 120 $\text{Br}_{bW t\varphi^0}$ & 0.05 & 0.13 \\
& 450 & $9.9 \times 10^{-4}$ & 88 $\text{Br}_{bW t\varphi^0}$ & 0.07 & 0.19
\end{tabular}
\caption{\label{tab:toppartnerneutralchannel} Efficiencies for passing the given selection criteria for several representative values of $m_{\varphi^0}$.  Also shown are corresponding values of $\epsilon$ yielding $2 \sigma$ and $5 \sigma$  significance assuming $\text{Br}(\varphi^0 \rightarrow b b) = 1$.  As in \Tab{tab:toppartnerdifferentmasses}, $2 \sigma$ ($5 \sigma$) significances correspond to $S \approx 5.5 \; (13.7)$ and $B \approx 7.7 \; (7.5)$.}
\end{table}

Efficiencies for two sample values of $m_{\varphi^0}$ are given in \Tab{tab:toppartnerneutralchannel}, along with corresponding values of $\epsilon$ yielding $2 \sigma$ and $5 \sigma$ significances (as above, taking $\text{Br}(T \rightarrow t \varphi^0) = \epsilon$ and $\text{Br}(T \rightarrow bW^\pm:th:tZ) = (1-\epsilon) \times (\frac{1}{2}:\frac{1}{4}:\frac{1}{4})$).  As expected, the efficiencies and branching ratios reach are comparable to the $T \rightarrow b H^\pm$ search.

\begin{figure}
\centering
\includegraphics[width=0.7\textwidth]{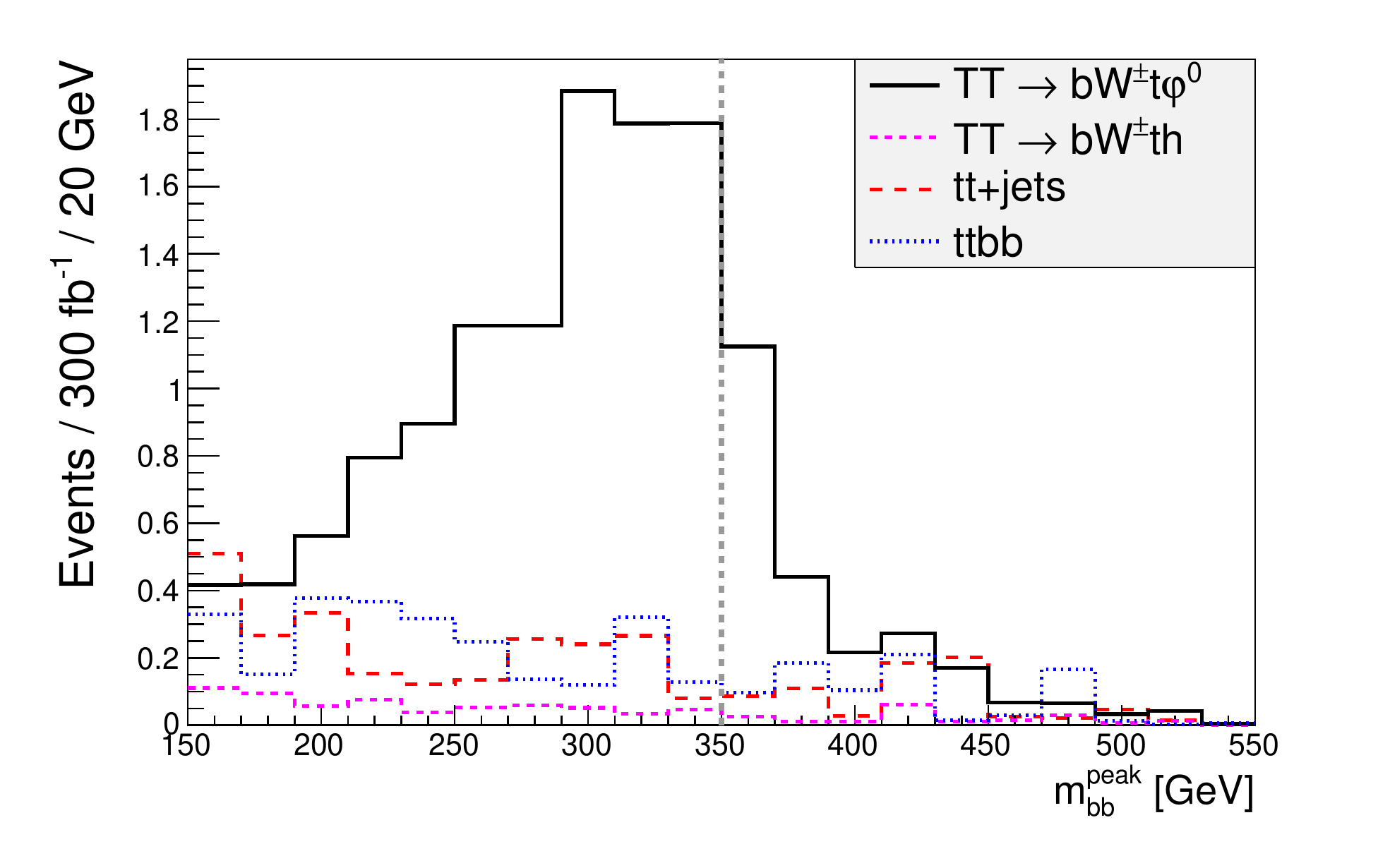}
\caption{Distribution of $m_{bb}^{\rm{peak}}$
taking $m_T = 700 \text{ GeV}, m_{\varphi^0} = 350 \text{ GeV}$, $\epsilon = 0.13$.  We assume $\text{Br}(\varphi^{0} \rightarrow bb) =1$. In contrast to \Fig{fig:mbbedgeplot}, this $b$ pair should reconstruct the $\varphi^0$, producing a resonance peak at $m_{\varphi^0}$ (dashed line).}
\label{fig:mbbpeakplot}
\end{figure}

Of course, the $bb$ pair produced in $T \rightarrow t \varphi^0 \rightarrow tbb$ should exhibit a resonance structure at $m_{bb} = m_{\varphi^0}$, so by employing a similar tactic to that used above to identify the edge (i.e.~by forming $m_{bb}^{\rm{peak}}$ using the pair of $b$'s in $\{b_2,b_3,b_4\}$ that do not give the minimum $|m_{b_k \ell \slashed{E}_T} - m_t|$) one could attempt to search for a resonance peak.  A sample distribution for $m_T = 700 \text{ GeV}, m_{\varphi^0} = 350 \text{ GeV}$, and $\epsilon = 0.13$ is shown in \Fig{fig:mbbpeakplot}.  The resonance peak is not particularly sharp in part because we are not using the full neutrino four-momentum to reject the $b$ jet from the top decay and mitigate combinatoric confusion.  The peak could potentially be improved by solving for the full four-momentum with $p_T^\nu = \slashed{p}_T$ and requiring $m_{\ell \nu} = m_W$ and $m_{bbb\ell\nu} = m_T$.
Again, the feasibility of discovering a resonance structure in this fashion is limited due to the small statistics, but such a structure could in principle help not only to determine $m_{\varphi^0}$ but also to distinguish between $T \rightarrow b H^\pm$ and $T \rightarrow t \varphi^0$.

\section{Conclusions}
\label{sec:Conclusion}
If the weak scale is in fact natural, new states should soon be discovered at the LHC.  These new states would of course provide insights into why the Higgs boson has a weak scale mass, but they might also provide an unexpected window into a rich scalar sector that would be otherwise difficult to access experimentally.  In this paper, we have argued that heavy charged Higgs bosons can be challenging to observe in standard channels, but they might well be discoverable in the decays of top partners.  Top partner decays can also be sensitive to exotic neutral scalars.

We have focused on methods for observing extended Higgs sector scalars that decay predominantly via $H^\pm \rightarrow t b$ or $\varphi^0 \rightarrow bb$.  These decay channels are likely to dominate if the extended Higgs sector scalars have large couplings to third-generation quarks.  That said, other decay modes may also be present depending on the exact structure of the theory.  For instance, decays like $H^\pm \rightarrow \tau^\pm \nu_\tau$ or $H^\pm \rightarrow W^\pm h$ may provide alternative signatures of scalars produced either directly or in fermionic top partner decays.

The strategy presented here makes use of the (likely significant) $T \rightarrow b W^\pm$ decay to tag top partner pair production events.  However, if other top partner decay modes dominate, alternative search strategies would be preferred.  In particular, if the top partner decays predominantly as $T \rightarrow th$, a cut on $\text{min}(m_{bb})$ can no longer be employed to separate signal from background.  The decay $TT \rightarrow thbH^\pm$ would yield a striking $6b$, $2W^\pm$ final state, but combinatoric backgrounds associated with the large number of $b$-jets would make it difficult to disentangle this decay pattern from, e.g., $TT \rightarrow thth$.  Similarly, bottom partners $B$ are also expected to be light if they are in an electroweak doublet with the top partner $T$, and the decay mode $BB \to t W^\pm t H^\pm$ ($BB \to b h t H^\pm$) yields a striking $4b$, $4W^\pm$ ($6b$, $2W^\pm$) final state, albeit with significant combinatoric confusion.

Finally, while this search strategy could reveal the presence of extended Higgs sector scalars, distinguishing between $T \rightarrow t \varphi^0 \rightarrow t b b$ and $T \rightarrow H^\pm b\rightarrow t b b$ would likely prove challenging given the small statistics.  Of course, the first priority is to determine the presence of additional scalar states, but how to determine their properties is a question of great interest, especially given the difficulty in uncovering them in the first place.  We leave these questions for future investigation, as we await hints of naturalness from the LHC.

\begin{acknowledgments}
We thank Timothy Cohen, Bogdan Dobrescu, and Martin Schmaltz for useful conversations.
J.K. thanks Josh Gevirtz for assistance with \textsc{Root}.
The work of  J.K. and A.P. is supported in part by NSF Career Grant NSF-PHY-0743315, with A.P. receiving additional support from the U.S. Department of Energy (DoE) Grant \#DE-SC0007859.
J.T. is supported by the U.S. DoE under cooperative research agreement DE-FG02-05ER-41360 and under the DoE Early Career research program DE-FG02-11ER-41741.
\end{acknowledgments}

\bibliographystyle{JHEP}
\bibliography{TtoChargedH}
\end{document}